\documentstyle[preprint,aps,psfig]{revtex}

\def\lsim{\raise0.3ex\hbox{$\;<$\kern-0.75em\raise-1.1ex
\hbox{$\sim\;$}}}
\def\gsim{\raise0.3ex\hbox{$\;>$\kern-0.75em\raise-1.1ex
\hbox{$\sim\;$}}}
\begin{document}
\textheight = 23.cm
\topmargin = -1.4cm
%\baselineskip 7.6mm
%\tightenlines

\preprint{
\vbox{
\hbox{Fermilab-Pub-02/064-T} \vskip -0.2cm
\hbox{IFT-P.024/2002}\vskip -0.2cm
\hbox{hep-ph/0204171}\vskip -0.2cm
\hbox{\hfill April 10, 2002}
}}
\title{
CP and T Trajectory Diagrams for a Unified Graphical Representation of
Neutrino Oscillations
}

\author{\large{
Hisakazu Minakata$^1$,
Hiroshi Nunokawa$^2$,
Stephen Parke$^3$
}}

\address{
$^1$
\sl Department of Physics, Tokyo Metropolitan University \\[-0.3cm]
1-1 Minami-Osawa, Hachioji, Tokyo 192-0397, Japan\\[-0.3cm]
minakata@phys.metro-u.ac.jp
}
\address{
$^2$
\sl Instituto de F\'{\i}sica Te\'orica,
Universidade Estadual Paulista \\[-0.3cm]
Rua Pamplona 145, 01405-900 S\~ao Paulo, SP Brazil\\[-0.3cm]
nunokawa@ift.unesp.br
}
\address{
$^3$
\sl Theoretical Physics Department,
Fermi National Accelerator Laboratory \\[-0.3cm]
P.O.Box 500, Batavia, IL 60510, USA\\[-0.3cm]
parke@fnal.gov
}

\maketitle

\vspace{-1.0cm}
%%%%%%%%%%%%%%%%%%%%%%%%%%%%%%%%%%%%%%%%%%%%%%%%%%%%%%%%%%%%%%%%%
%    Abstract
%%%%%%%%%%%%%%%%%%%%%%%%%%%%%%%%%%%%%%%%%%%%%%%%%%%%%%%%%%%%%%%%%
\hfuzz=25pt
\begin{abstract}
Recently the CP trajectory diagram was introduced to demonstrate
the difference between the intrinsic CP violating effects to
those induced by matter for neutrino oscillation.
In this paper we introduce the T trajectory diagram. 
In these diagrams the probability for a given oscillation process
is plotted versus the probability for the CP- or the T-conjugate 
processes, which forms an ellipse as the CP or T violating phase 
is varied. 
Since the CP and the T conjugate processes are related 
by CPT symmetry, even in the presence of matter, these two trajectory 
diagrams are closely related with each other and form a unified 
description of neutrino oscillations in matter.

\end{abstract}

%\vskip2pc]

\newpage
%%%%%%%%%%%%%%%%%%%%%%%%

\section{Introduction}

Accumulating evidences for neutrino oscillation in the
atmospheric \cite{SKatm}, the solar \cite{solar}, and the accelerator
\cite{K2K} neutrino experiments make it realistic to think about
exploring the full structure of the lepton flavor mixing.
One of the challenging goals in such attempt would be to measure
the leptonic CP or T violating phase $\delta$ in the MNS matrix
\cite{MNS} \footnote{In this paper we will assume that the light 
neutrino sector consists of only three active neutrinos.}.
It appears that the long baseline neutrino oscillation experiments
are the most feasible way to actually detect these effects.

It has been known since sometime ago that the measurement of
CP and T violating effects in the long baseline neutrino oscillation
experiments can be either contaminated or enhanced by the matter 
effect inside the earth; see e.g., Refs.~\cite{cp-matter} 
and \cite{t-matter}.
Therefore, it is one of the most important issues to achieve 
a complete understanding of the interplay between the CP-phase 
and the matter effects in parameter regions relevant for 
such experiments.

Toward the goal, we have introduced in a previous paper
\cite{MNjhep01} the
``CP trajectory diagram in bi-probability space'' as a useful
tool for pictorial representation of the CP and the matter
effects in neutrino oscillation. This diagram enables us to display
three effects;
(a) genuine CP violation due to the $\sin \delta$ term,
(b) CP conserving $\cos \delta$ term, and
(c) fake CP violation due to the earth's matter,
separately in a single diagram.

In this paper, we introduce a related but an entirely new diagram,
``T trajectory diagram in bi-probability space''. We demonstrate
that the T trajectory diagram is by itself illuminative and is
complementary with the CP diagram. More importantly, when
combined with the CP diagram, it completes
and unifies our understanding of the interplay between the
effects due to CP or T-violating phase and the matter.
The intimate relationship between the T and
the CP diagrams reflects the underlying CPT theorem.
Although the CPT theorem itself is broken by the presence of matter,
this breaking can be compensated by a change 
in the sign of all the $\Delta m^2$'s.

\section{CP and T trajectory diagrams in bi-probability space and
their symmetries}

Now we introduce the T trajectory diagram in
bi-probability space spanned by
$P(\nu) \equiv P(\nu_{e} \rightarrow \nu_{\mu})$ and
its T-conjugate
T$[P(\nu)] \equiv P(\nu_{\mu} \rightarrow \nu_{e})$.
We fully explain in this section its notable characteristic
properties, the relationship (or unity) with the CP
trajectory diagram, and the symmetry relations obeyed by them.

Suppose that we compute the oscillation probability $P(\nu)$ and
T$[P(\nu)]$ with a given set of oscillation and experimental
parameters. Then, we draw a dot on the two-dimensional plane spanned
by $P(\nu)$ and T$[P(\nu)]$.
When $\delta$ is varied we have a set of dots which forms a closed
trajectory, closed because the probability must be a periodic
function of $\delta$, a phase variable.
Let us remind the reader that the T trajectory diagram has very
similar structure with the CP trajectory diagram introduced
in Ref.~\cite{MNjhep01}; the abscissa is the same and the ordinate
for the CP diagram is 
CP$[P(\nu)] \equiv P(\bar{\nu}_{e} \rightarrow \bar{\nu}_{\mu})$.

In Fig. 1 the CP and the T trajectory diagrams, denoted as 
CP$\pm$ and T$\pm$, are plotted in
the same figure. Here, the ordinate of the diagram is meant to be
CP$[P(\nu)]$ for the CP and T$[P(\nu)]$ for the T diagrams,
respectively.
In the center there exist two vacuum diagrams, V$\pm$, 
one for each of the signs of 
$\Delta m^2_{31}$.
When the matter effect is turned on, the positive and the
negative $\Delta m^2_{31}$ trajectories split.

In vacuum, the CP and the T trajectories are identical with 
each other. This is because the CPT theorem tells us that
T [$P(\nu)$] $\equiv P(\nu_{\mu} \rightarrow \nu_{e}) =
P(\bar{\nu}_{e} \rightarrow \bar{\nu}_{\mu})$ = CP$[P(\nu)]$.
Depending upon the sign of $\Delta m^2_{31}$ we have two CP (or T)
trajectories which are almost degenerate with each other in vacuum
due to the approximate symmetry under the simultaneous transformation
\cite{MNjhep01}
\begin{eqnarray}
& &\delta \rightarrow \pi - \delta
\hskip 0.5 cm (\mbox{mod.} 2 \pi),
\nonumber \\
& &\Delta m^2_{31} \rightarrow - \Delta m^2_{31}.
\label{flipsym}
\end{eqnarray}

It was noticed that the CP and 
the T trajectories are elliptical exactly in vacuum and 
approximately in matter \cite{MNjhep01}.
Recently, it was shown in a remarkable paper 
by Kimura, Takamura, and Yokomakura \cite{KTY02} that 
it is exactly elliptic even in matter;
the $\delta$-dependence of the oscillation probabilities in 
constant matter density can be written on general ground in the form 
\begin{eqnarray}
P(\nu) &=& A(a) \cos{\delta} + B(a) \sin{\delta} + C(a) 
\nonumber \\
CP[P(\nu)] &=& A(-a) \cos{\delta} - B(-a) \sin{\delta} + C(-a) 
\label{general} \\
T[P(\nu)] &=& A(a) \cos{\delta} - B(a) \sin{\delta} + C(a) \nonumber 
\end{eqnarray}
where $a$ denotes neutrino's index of refraction in matter, 
$a = \sqrt{2} G_F N_e$, with electron number density $N_e$ 
and the Fermi constant $G_F$.
The dependence on other variables are suppressed.
Throughout this paper, we use the standard parameterization 
of the MNS matrix.

Matter effects split the two trajectories in quite different 
manners. Namely, the T trajectories split along
the straight line $P(\nu)=$ T$[P(\nu)]$
whereas the CP trajectories split in the orthogonal 
direction.\footnote{
%%%%%%%%%%%%%  Footnote 2  %%%%%%%%%%%%%%%%%%%%%%%%%%%%%%%%%%%%%%%
While it may be easier for the readers to understand the 
following discussions by referring the KTY formula in Eq. (\ref{general}), 
our subsequent discussion will be entirely independent from the 
constant density approximation. We however rely on the adiabatic 
approximation when we utilize perturbative formulas.}
The T trajectories must move along the diagonal line because
the $\delta = 0, ~\pi$ points on T trajectories must stay on the diagonal
because T violation vanishes at $\delta = 0, ~\pi$ for matter distributions
which are symmetric about the mid-point between production and
detection.
In constant matter density, this stems from 
the Naumov-Harrison-Scott relation 
(i.e., the proportionality between the vacuum and 
the matter Jarlskog factors) \cite{N92,HS00} 
and the absence of $\sin{\delta}$ dependence \cite {ZS88} 
in the other parts of the oscillation probability, 
or simply from the KTY formula in Eq. (\ref{general}).

On the other hand, the CP trajectories must split along the
direction orthogonal to the diagonal line for small matter effect.
This is because the first order matter correction, which is
proportional to $a = \sqrt{2} G_F N_e(x)$, has opposite sign
for the neutrino and the anti-neutrino oscillation probabilities;
$\Delta P(\nu) = - \Delta P(\bar{\nu})$ \cite{MNjhep01}.
No matter how large the matter effect the line connecting
the two CP trajectories with opposite signs of $\Delta m^2_{31}$
is orthogonal to the diagonal line. As we will explain
below this reflects a symmetry relationship obeyed by
the oscillation probabilities, which we denote as the
``CP-CP relation''.

One may also notice, as indicated by the eye-guided lines in
Fig. 1 that the CP and the T trajectories have identical lengths
when projected onto either the abscissa or the ordinate.
It should be the case for the projection onto the abscissa
because the abscissa is common for both of the CP and the
T diagrams. What is nontrivial is the equality in length
when projected onto the ordinate. It again represents a
symmetry which we want to call the ``T-CP relation''.
We note that neither the CP-CP nor the T-CP relations
are exact, as one may observe from Fig.1.

The precise statement of the CP-CP relation is (see Fig.~1)
\begin{eqnarray}
P(\nu_{e} \rightarrow \nu_{\mu};
\Delta m^2_{31},\Delta m^2_{21}, \delta,a)
&=&
P(\bar{\nu}_{e} \rightarrow \bar{\nu}_{\mu};
-\Delta m^2_{31},-\Delta m^2_{21},\delta, a) \nonumber \\
& \approx &
P(\bar{\nu}_{e} \rightarrow \bar{\nu}_{\mu};
-\Delta m^2_{31},+\Delta m^2_{21},\pi + \delta, a). 
\label{CP-CP}
\end{eqnarray}
Whereas the precise statement of the T-CP relation is
\begin{eqnarray}
P(\nu_{\mu} \rightarrow \nu_{e};
\Delta m^2_{31},\Delta m^2_{21}, \delta, a)
&=&
P(\bar{\nu}_{e} \rightarrow \bar{\nu}_{\mu};
-\Delta m^2_{31},-\Delta m^2_{21},2 \pi - \delta, a) \nonumber  \\
& \approx &
P(\bar{\nu}_{e} \rightarrow \bar{\nu}_{\mu};
-\Delta m^2_{31},+\Delta m^2_{21}, \pi - \delta, a ).
\label{T-CP}
\end{eqnarray}
The equalities have the opposite sign for all the $\Delta m^2$'s
from the left hand side to the right hand side.
But because of the hierarchy,
$|\Delta m^2_{31}| \gg |\Delta m^2_{21} |$, changing the sign of 
$\Delta m^2_{21}$ produces only a small deviation whose precise 
origin and magnitude will be become clear in the derivation of these
relations. 
These relations imply that when the
CP+ trajectory winds counter-clockwise as $\delta$ increases
the CP- and T- (T+) trajectories wind clockwise (counter-clockwise)
as in Fig.1.

To derive the CP-CP and T-CP relationships we need a number of identities.
The first set of identities comes from taking the time reversal and
the complex conjugate of the neutrino evolution equation assuming 
that the matter profile is symmetric about the mid-point
between production and detection;
\begin{eqnarray}
P(\nu_{\alpha} \rightarrow \nu_{\beta};
\Delta m^2_{31},\Delta m^2_{21}, \delta, a)
&=&
P(\bar{\nu}_{\beta} \rightarrow \bar{\nu}_{\alpha};
\Delta m^2_{31},\Delta m^2_{21}, \delta, -a).
\label{CPT-ids}
\end{eqnarray}
These identities are just CPT invariance in the presence of matter.

The second set of identities comes from taking the complex 
conjugate of the neutrino evolution equation
for an arbitrary matter distribution, they are
\begin{eqnarray}
&P(\nu_{\alpha} \rightarrow \nu_{\beta};
~\Delta m^2_{31},~\Delta m^2_{21}, \delta,a)
&=
P(\nu_{\alpha} \rightarrow \nu_{\beta};-\Delta m^2_{31},-\Delta m^2_{21},
2 \pi - \delta, -a) \nonumber \\
=
&P(\bar{\nu}_{\alpha} \rightarrow \bar{\nu}_{\beta};
-\Delta m^2_{31},-\Delta m^2_{21},\delta, a) \quad
&=
P(\bar{\nu}_{\alpha} \rightarrow \bar{\nu}_{\beta};
~\Delta m^2_{31},~\Delta m^2_{21},2 \pi - \delta, -a)
\label{CP-ids}
\end{eqnarray}
These identities relate the (anti-) neutrino oscillation probabilities
in matter to those in anti-matter with opposite signs for all
the $\Delta m^2$'s.
They also relate the neutrino oscillation probabilities in matter
to the anti-neutrino oscillation probabilities in matter with 
opposite signs for all the $\Delta m^2$.

Combinations of these identities give the equalities in the CP-CP 
and the T-CP relations, the first lines in 
Eqs. (\ref{CP-CP}) and (\ref{T-CP}).
Now we will use the hierarchy that
$|\Delta m^2_{31}| \gg |\Delta m^2_{21}|$ to get an approximate
equality if we flip the sign of the $\Delta m^2_{21}$ with
the appropriate change in the phase $\delta$.
The probability for $\nu_{\alpha} \rightarrow \nu_{\beta}$
consists of three terms as in Eq. (\ref{general}).
The coefficients $A$ and $B$ vanish like 
$\Delta m^2_{21}$ as $\Delta m^2_{21} \rightarrow 0$.
Thus, a change in the sign of $\Delta m^2_{21}$ in these two 
coefficients can be compensated by replacing $\delta$ with 
$\pi+\delta$, whereas the change in the coefficients $C$ is 
further suppressed by an extra factor of $\sin \theta_{13}$.
This proves the approximate equalities in the CP-CP and the 
T-CP relations. 

The CP-CP relation guarantees that the size and the shape of 
two sign-conjugate ($\Delta m^2_{31}=\pm |\Delta m^2_{31}|$) 
diagrams are identical, 
whereas there is no such relation in T diagrams. 
The CP phase relation between two sign-conjugate CP diagrams implies, 
among other things, that the two-fold ambiguity in $\delta$ 
which still remains after accurate determination of $\theta_{13}$ 
\cite {MNjhep01} are related by 
$\delta_{2} = \delta_{1} + \pi$
apart from the correction of order 
$\sim \frac{\Delta m^2_{21}}{\Delta m^2_{31}}$.

Finally, we note that from the first line of Eq. (\ref{CP-ids}) that 
\begin{equation}
P(\nu_{\alpha} \rightarrow \nu_{\beta};
~\Delta m^2_{31},~\Delta m^2_{21}, \delta,a)
\approx
P(\nu_{\alpha} \rightarrow \nu_{\beta};-\Delta m^2_{31},\Delta m^2_{21},
\pi - \delta, -a) 
\label{flipsym2}
\end{equation}
which is a generalization of the approximate symmetry under 
the transformation (\ref{flipsym}) into the case in matter, 
from which the CP-CP and the T-CP relations also follow.

\section{Further Examples}

\begin{figure}[h]
\vspace*{14cm}
\includegraphics{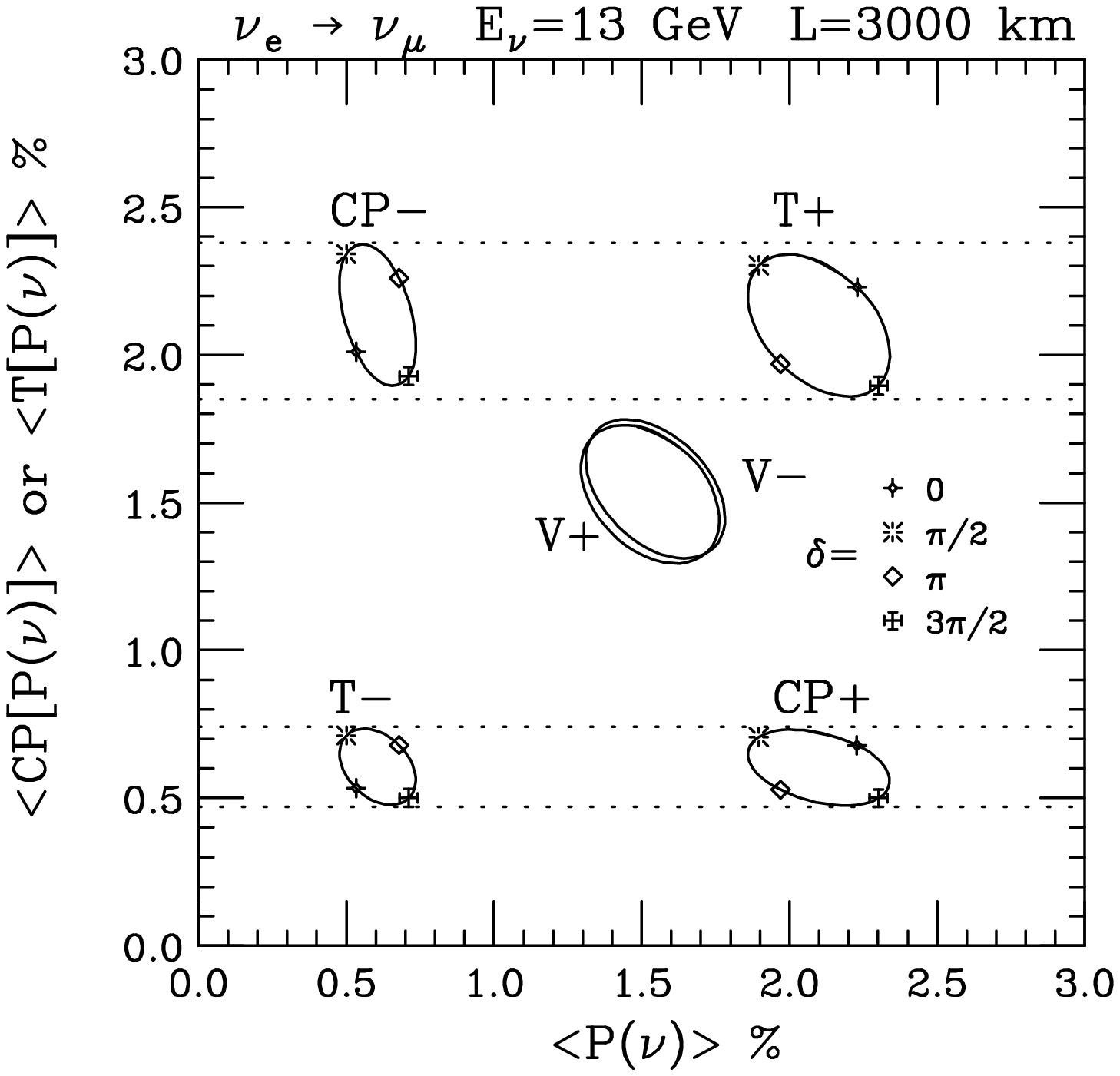}
\vspace{1.0cm}
\caption[]{
The T (CP) trajectory diagrams (ellipses) in the plane
$P(\nu_e \rightarrow \nu_\mu)$ verses $P(\nu_\mu \rightarrow \nu_e)$
($P(\bar{\nu}_e \rightarrow \bar{\nu}_\mu)$) for an average
neutrino energy of 13 GeV, spread 20\%, and baseline of 3000 km.  
The ellipses labelled with a T and CP are in matter with
a density times electron fraction given by $ Y_e \rho  = 1.5$ g cm$^{-3}$
whereas those ellipses labelled V are in vacuum where the T and CP
trajectories are identical.
The plus or minus indicates the sign of $\Delta m^2_{31}$.
The mixing parameters are fixed to be 
$|\Delta m^2_{31}| = 3 \times 10^{-3} eV^2$,
$\sin^2 2\theta_{23}=1.0$,
$\Delta m^2_{21} = +5 \times 10^{-5} eV^2$,
$\sin^2 2\theta_{12}=0.8$ and
$\sin^2 2\theta_{13}=0.05$.
The dotted lines are to guide the eye in demonstrating the T-CP
relationship given by eqn(3).
The marks on the CP and T ellipses are the points where 
the CP or T violating phase 
$\delta = (0,1,2,3){\pi/2}$ as indicated.
}
\label{e13fig}
\end{figure}

\begin{figure}[h]
\vspace*{7.7cm}
\includegraphics{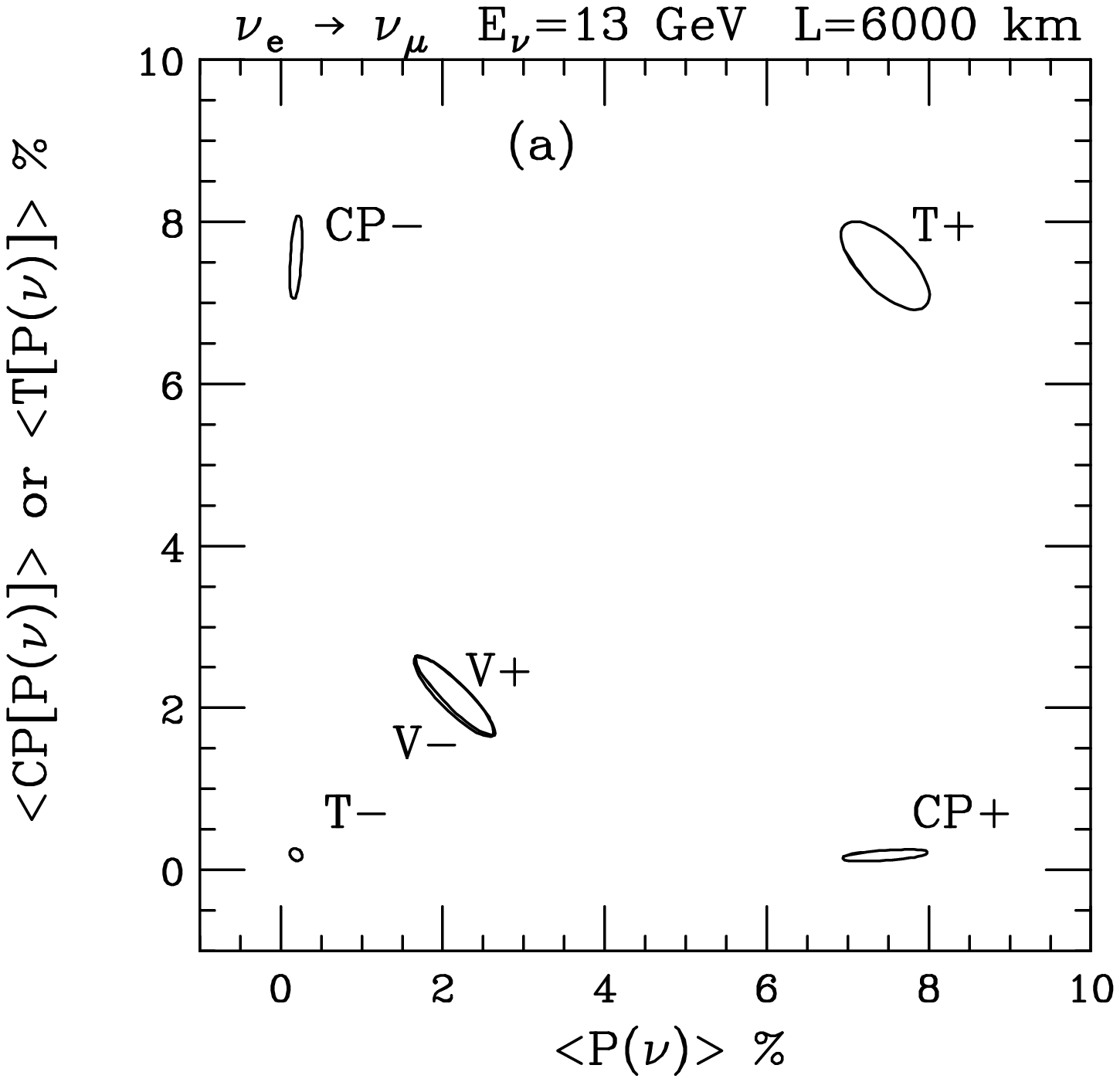}
\includegraphics{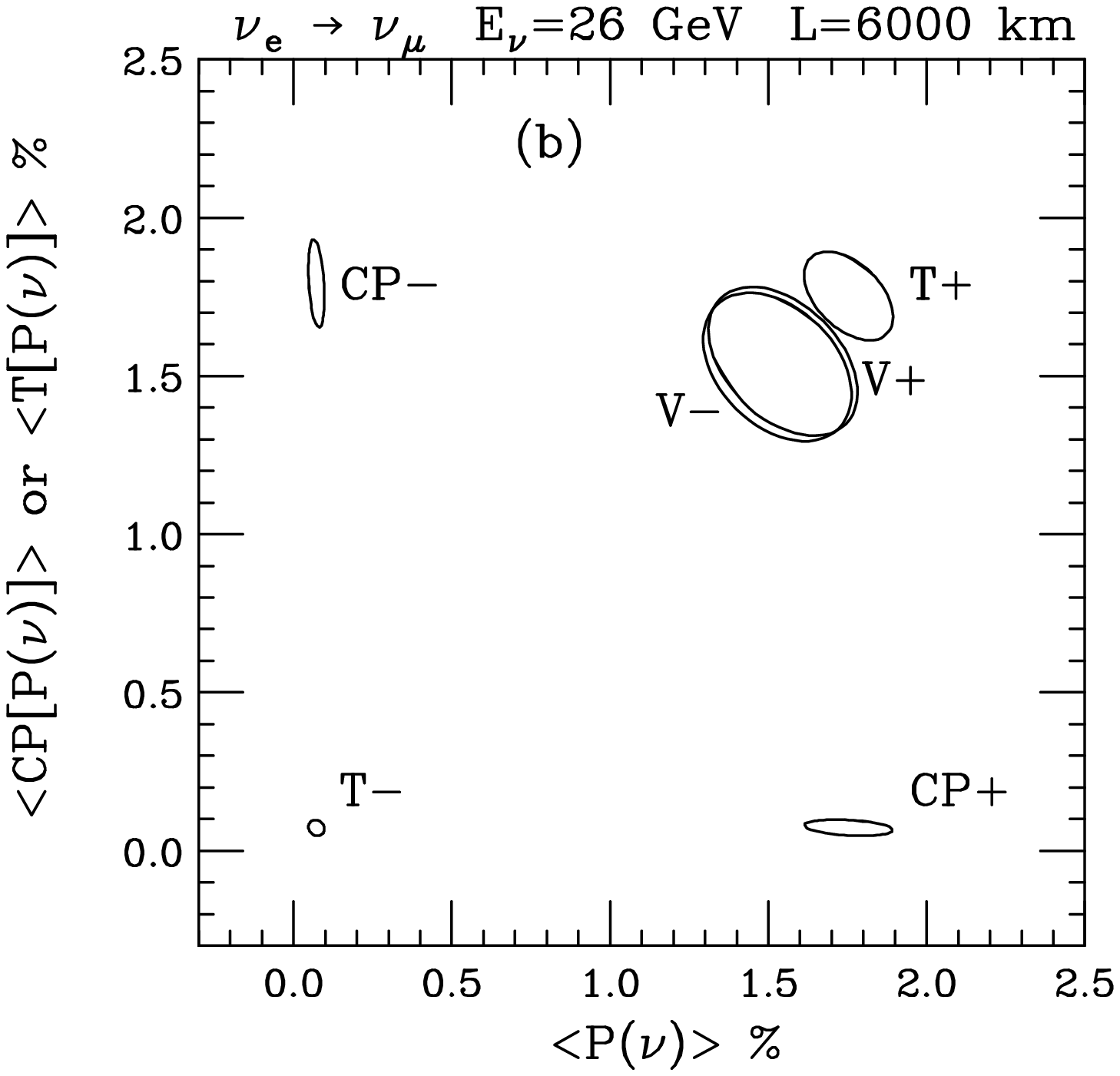}
\vspace{1.0cm}
\caption[]{
The T (CP) trajectory diagrams (ellipses) in the plane
$P(\nu_e \rightarrow \nu_\mu)$ verses $P(\nu_\mu \rightarrow \nu_e)$
($P(\bar{\nu}_e \rightarrow \bar{\nu}_\mu)$) for an average
neutrino energy (spread 20\%) and baseline of (a) 13 GeV and 6000 km 
and (b) 26 GeV and 6000 km. Labels and mixing parameters are as in Fig. 1.}
\label{L6000fig}
\end{figure}

While Fig. \ref{e13fig} is for the preferred parameters of a neutrino factory
\cite{nufact} it is interesting to see how the T-CP trajectory
diagram changes as we change the energy and path length of the
experiment.  In Fig. \ref{L6000fig} we have increased the path length
to 6000 km and given the CP-T diagram for neutrino energies of
13 and 26 GeV.
Again one can see how well the CP-CP and T-CP relationships hold.

The next examples use an energy which is approximately half the
resonance energy corresponding to $\Delta m^2_{31}$ as suggested by
Parke and Weiler \cite{t-matter}. 
At a distance of 3500 km this energy
maximizes T-violation effects which is reflected in the size of ellipses. 
Halving the distance between source and detector for this energy
produces a T-CP trajectory diagram which is similar to the
neutrino factory diagram, Fig. \ref{e13fig} but with larger probabilities
and asymmetries.

The final examples are using the energies and baselines of 
NUMI/MINOS and JHF/SK see Fig. \ref{e1fig}.
For NUMI/MINOS there is reasonable separation between the two
$\Delta m^2_{31}$ ellipse whereas for JHF/SK there is some overlap.
From the viewpoint of simultaneous determination of $\delta$ 
and the sign of $\Delta m^2_{31}$ the longer baseline of 
NUMI/MINOS would be more advantageous, 
while there are possibilities that it can be done at JHF/SK 
if $P$ and CP[$P$] are asymmetric \cite{MNjhep01,taup2001},

An another feature which is worth noting in the JHF/SK case 
in Fig. \ref{e1fig} is that the problem of parameter degeneracy 
in T violation measurements is milder than that in CP 
measurements. It is because the T (or CP) ellipse is flatter 
in the radial direction, i.e., along the movement of T trajectory 
due to matter effect. 
The underlying reason for the phenomenon is that the coefficient 
of the $\cos{\delta}$ term oscillates and hence is always 
smaller than the coefficient of the $\sin{\delta}$ term 
when averaged over energy width of a beam. In this sense, 
T measurement, if feasible experimentally, is more advantageous 
than CP measurement for simultaneous determination of 
$\delta$ and the sign of $\Delta m^2_{31}$.

\section{Parameter degeneracy in T-violation measurements}

Armed by the T as well as the CP trajectory diagrams we are now 
ready to discuss the problem of parameter degeneracy with 
T-violation measurement. 
We do not aim at complete treatment of the problem in this paper but 
briefly note the new features that arise in T-violation measurement 
as compared to the CP-violation measurement. They arise 
because of the more symmetric nature of the T conjugate probability 
as apparent in Eq. (\ref{general}).

We work in the same approximation as in Ref.~\cite{BurguetC}  
and write the oscillation probability in small $s_{13 }$ 
approximation.  We have four equations: 
\begin{eqnarray}
P(\nu)_{\pm} &=& X_{\pm} \theta^2 + 
Y_{\pm} \theta \cos {\left( \delta \mp \frac{\Delta_{31}}{2} \right)} + 
P_{\odot}
\nonumber \\
T[P(\nu)]_{\pm} &=& X_{\pm} \theta^2 + 
Y_{\pm} \theta \cos {\left( \delta \pm \frac{\Delta_{31}}{2} \right)} + 
P_{\odot} 
\label{Tequation}
\end{eqnarray}
where $X_{\pm}$ and $Y_{\pm}$ are given in Ref.~\cite{BurguetC}, 
$P_{\odot}$ indicates the term which is related with 
solar neutrino oscillations, and 
$\Delta_{31} \equiv \frac{|\Delta m^2_{31}| L}{2E}$. 
Note that $\pm$ here refers to the sign of $\Delta m^2_{31}$ 
and $\theta$ is an abbreviation of $\theta_{13} \simeq s_{13}$.

It is easy to show that Eq. (\ref{Tequation}) can be solved 
to obtain the same-sign $\Delta m^2_{31}$ degenerate solutions as
\begin{equation}
\delta_{2} = \pi - \delta_{1} 
\hskip 0.5cm \mbox{and} \hskip 0.5cm 
\theta_{2} -  \theta_{1} = 
\frac{Y_{\pm}}{X_{\pm}} 
\cos{\delta_{1}}
\cos{\left(\frac{\Delta_{31}}{2}\right)}.
\end{equation}
Notice that it is the degeneracy in matter though it looks 
like the one in vacuum \cite{MNjhep01}.

Two T$\pm$ trajectories may overlap for a distance shorter than 
those in Fig.\ref{e1fig}, which would result in a mixed-sign degeneracy 
as in the case of CP measurement mentioned before. 
However, one can show that in the case of T violation measurement 
there is no more ambiguity in 
$\delta$ once $\theta_{13}$ is given; 
\begin{equation}
\delta_{2} = \delta_{1} 
\hskip 0.5cm \mbox{and} \hskip 0.5cm 
\cos{\delta_{1}} = - \theta_{13} 
\frac{(X_{+} - X_{-})}{2 Y_{+}\cos{\left(\frac{\Delta_{31}}{2}\right)}}
\end{equation}
apart from the computable higher order correction due to the 
matter effect. 
This is obvious from Fig.\ref{e1fig}(b). 
A fuller treatment of the parameter degeneracy with T violation 
measurement will be reported elsewhere \cite{MNP02}.

\section{Discussion and Conclusions}

First, it is worth pointing out that in matter the sensitivity to the CP
or T-violating phase, $\delta$, is larger for the T-violating
pair 
($\nu_\alpha \rightarrow \nu_\beta$, $\nu_\beta \rightarrow \nu_\alpha$)
than for the CP-violating pair 
($\nu_\alpha \rightarrow \nu_\beta$, $\bar{\nu}_\alpha \rightarrow 
\bar{\nu}_\beta$)
of processes assuming $\Delta m^2_{31} > 0$. 
For $\Delta m^2_{31} < 0$ one should use the following
T-violating pair,
($\bar{\nu}_\alpha \rightarrow \bar{\nu}_\beta$, $\bar{\nu}_\beta \rightarrow 
\bar{\nu}_\alpha$).
This can be seen by comparing the size of the T ellipses to
that of the CP ellipses in all of the previous diagrams.
Unfortunately the experimental challenges associated with performing
a T-violating experiment have yet to be overcome.

Due to the existence of the T-CP relationships for neutrino oscillations
it is instructive and useful to add the T-violating trajectories to
the CP-violating trajectories first proposed by Minakata and Nunokawa.
The size and position of the T-violating ellipses can be easily estimated
using simple arguments including the effects of matter and from them 
the CP-violating ellipses can be estimated.
Thus, the combination of the CP and T trajectories form a unified
picture of the CP-matter interplay 
in neutrino oscillations.

%%%%%%%%%%%%%%%%
\acknowledgments
%%%%%%%%%%%%%%%%

HM thanks Theoretical Physics Department of Fermilab for warm
hospitality extended to him during two visits when this work
has been initiated and then further pursuit.
He is grateful to Center for Theoretical Physics of MIT for
hospitality and support.
This work was supported by the Grant-in-Aid for Scientific Research
in Priority Areas No. 12047222, Japan Ministry
of Education, Culture, Sports, Science, and Technology.
HN was also supported by the Brazilian FAPESP Foundation.

SP would like to thank the Physics Department at the University of Auckland,
New Zealand were part of this work was performed.
Fermilab is operated by URA under DOE contract No.~DE-AC02-76CH03000.

%%%%%%%%%%%%%%%%%%%%%%%%%% Bibliography %%%%%%%%%%%%%%%%%%%%%%%%%%%%%%

\newpage 

\begin{figure}[h]
\vspace*{7.7cm}
\includegraphics{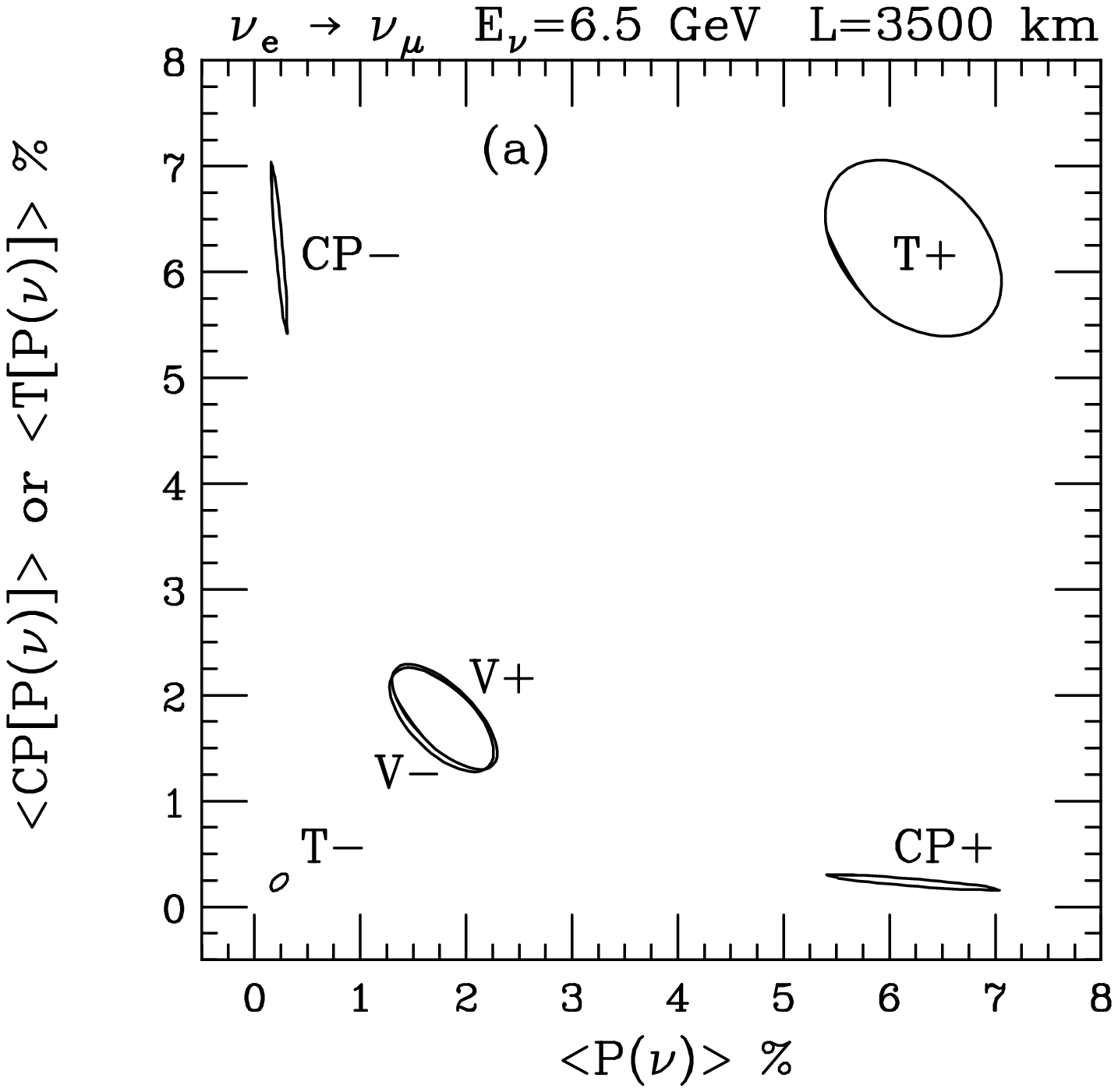}
\includegraphics{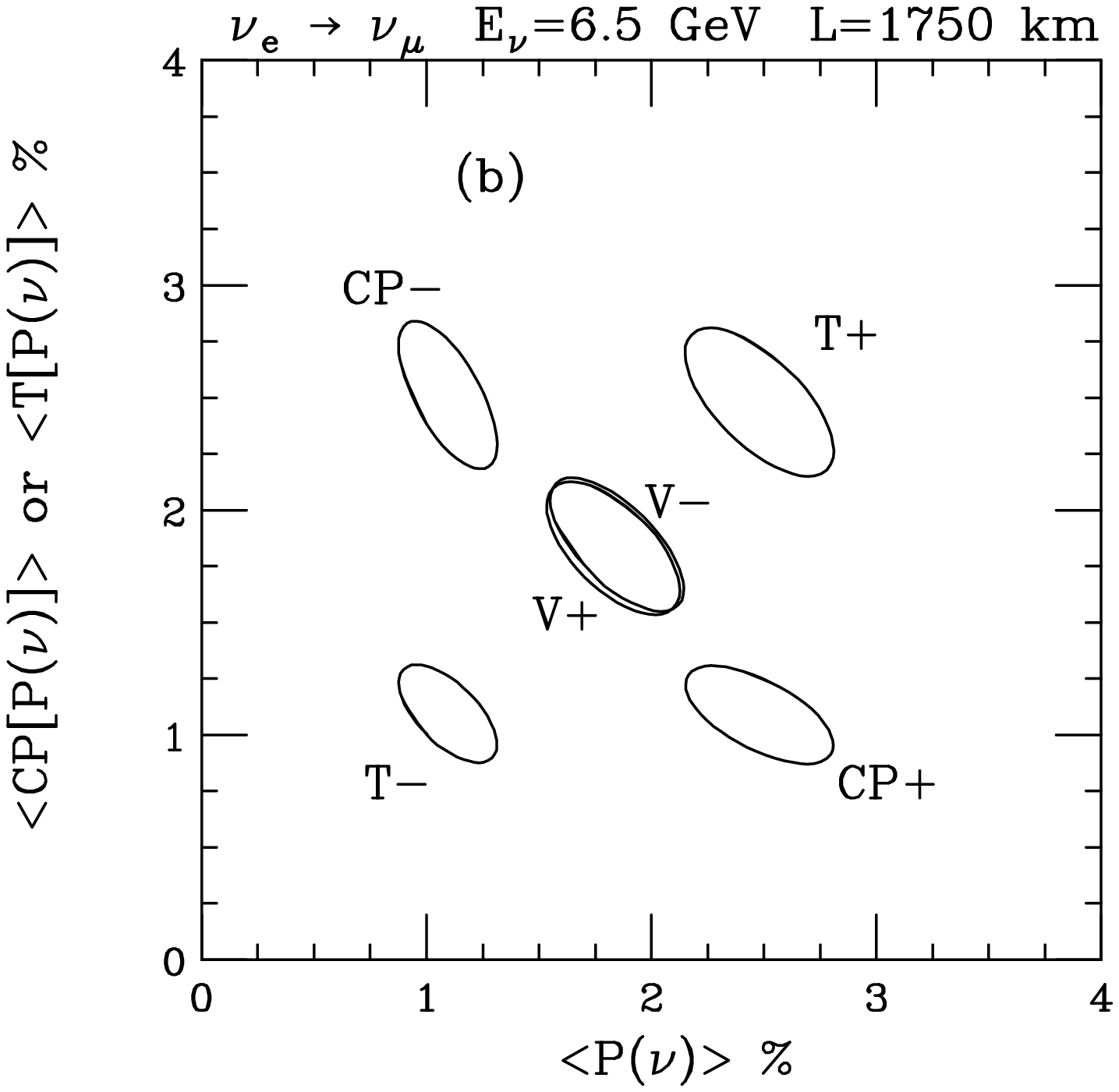}
\vspace{1.0cm} 
\caption[]{
The T (CP) trajectory diagrams (ellipses) in the plane
$P(\nu_e \rightarrow \nu_\mu)$ verses $P(\nu_\mu \rightarrow \nu_e)$
($P(\bar{\nu}_e \rightarrow \bar{\nu}_\mu)$) for an average
neutrino energy (spread 20\%) and baseline of (a) 6.5 GeV and 3500 km 
and (b) 6.5 GeV and 1750 km. Labels and mixing parameters are as in Fig. 1.}
\label{e6pt5fig}
\end{figure}

\newpage 

\begin{figure}[h]
\vspace*{8cm}
\includegraphics{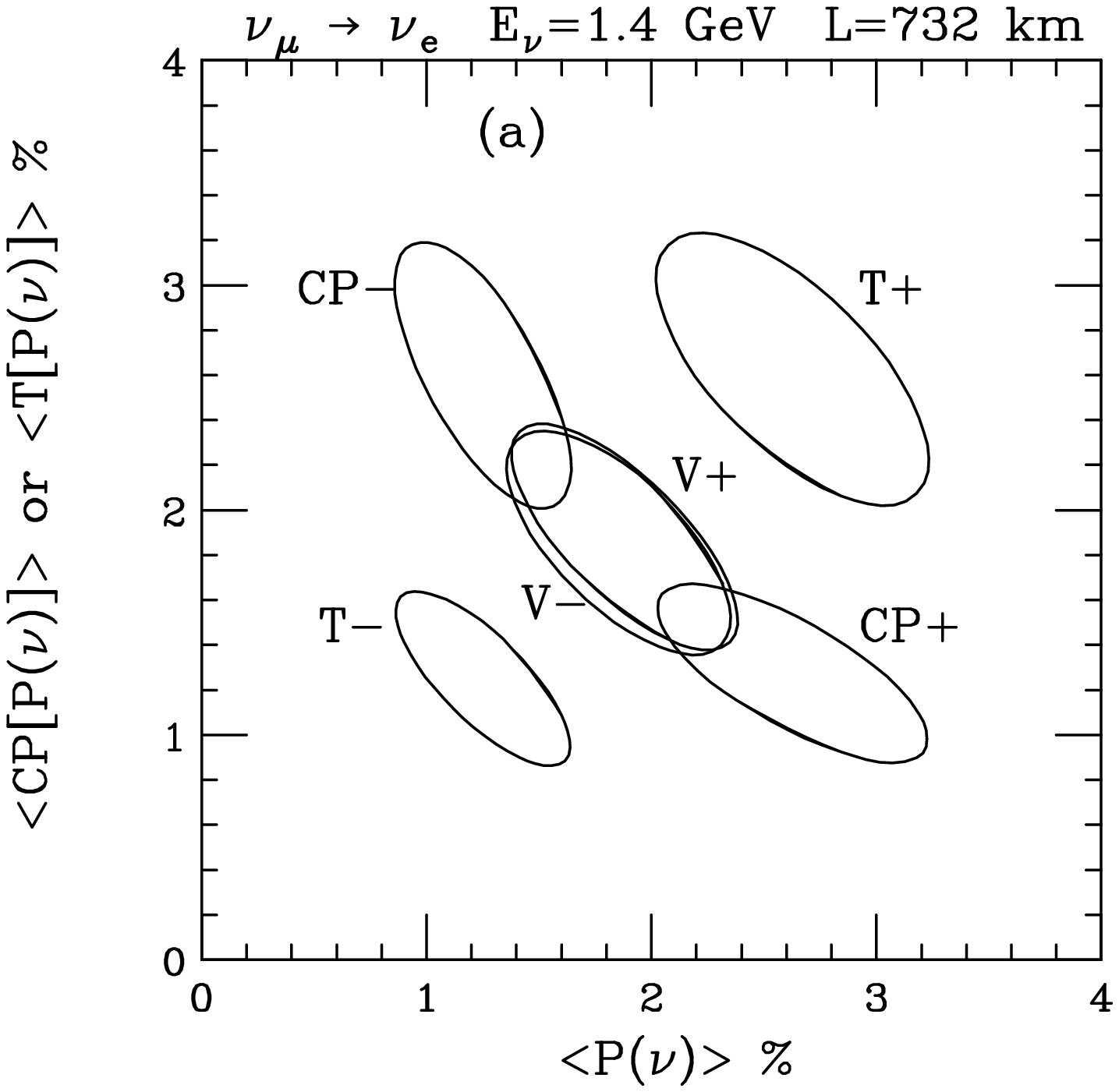}
\includegraphics{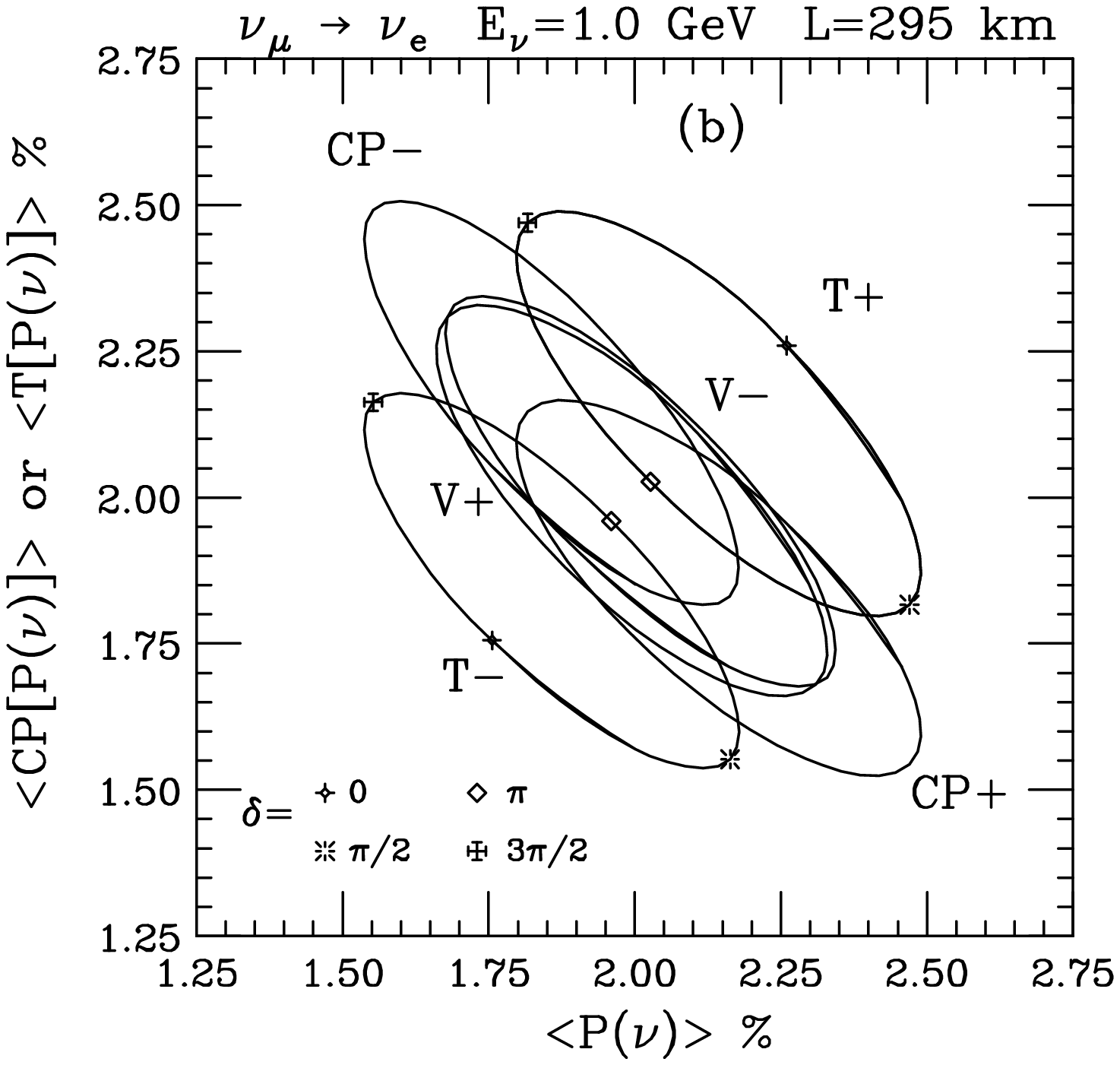}
\vspace{1.0cm} 
\caption[]{
The T (CP) trajectory diagrams (ellipses) in the plane
$P(\nu_\mu \rightarrow \nu_e)$ verses $P(\nu_e \rightarrow \nu_\mu)$
($P(\bar{\nu}_\mu \rightarrow \bar{\nu}_e)$) for an average
neutrino energy (spread 20\%) and baseline of (a) 1.4 GeV and 732 km 
(NUMI/MINOS) and (b) 1.0 GeV and 295 km (JHF/SK). 
Labels and mixing parameters are as in Fig. 1.}
\label{e1fig}
\end{figure}


\begin{thebibliography}{99}

\bibitem {SKatm}
Kamiokande Collaboration, Y. Fukuda {\it et al.},
Phys. Lett. B{\bf 335} (1994) 237;\\
Super-Kamiokande Collaboration, Y. Fukuda {\it et al.},
Phys. Rev. Lett. {\bf 81} (1998) 1562;
S. Fukuda {\it et al.}, {\it ibid.} {\bf 85} (2000) 3999.

\bibitem {solar}
Homestake Collaboration, B. T. Cleveland {\it et al.},
Astrophys. J.\ {\bf 496} (1998) 505;\\
%
SAGE Collaboration, J.\ N.\ Abdurashitov {\it et al.},
Phys.\ Rev.\ C {\bf 60} (1999) 055801; \\
%
GALLEX Collaboration, W.\ Hampel {\it et al.}, Phys.\
Lett.\  B {\bf 447} (1999) 127; \\
%
GNO Collaboration, M.~Altmann {\it et al.}  
%``GNO solar neutrino observations: Results for GNO I,''
Phys.\ Lett.\ B {\bf 490} (2000) 16; \\
%
Super-Kamiokande Collaboration,  S.\ Fukuda {\it et al.},
Phys. Rev. Lett. {\bf 86} (2001) 5651;
{\it ibid.}  {\bf 86} (2001) 5656;\\
%
SNO Collaboration, Q. R. Ahmad {\it et al.},
Phys. Rev. Lett. {\bf 87} (2001) 071301.

\bibitem {K2K}
K2K Collaboration, S.~H.~Ahn {\it et al.},
Phys.\ Lett.\ B {\bf 511} (2001) 178;\\
%%CITATION = HEP-EX 0103001;%%
See also http://neutrino.kek.jp/news/2001.07.10.News/index-e.html.


\bibitem {MNS}
Z.~Maki, M.~Nakagawa and S.~Sakata,
Prog.\ Theor.\ Phys.\  {\bf 28} (1962) 870.

\bibitem {cp-matter}
J. Arafune and J. Sato, Phys. Rev. D {\bf 55} (1997) 1653;
J. Arafune, M. Koike and J. Sato, Phys. Rev. D {\bf 56} (1997) 3093
[Erratum {\it ibid.} D {\bf 60} (1999) 119905]; \\
H. Minakata and H. Nunokawa, Phys. Rev. D {\bf 57} (1998) 4403;
Phys. Lett. B {\bf 413} (1997) 369; Phys. Lett. B {\bf 495} (2000) 369;
Nucl. Instrum. Meth. A {\bf 472} (2001) 421; \\
K.~Dick, M.~Freund, M.~Lindner and A.~Romanino,
Nucl.\ Phys.\ B {\bf 562} (1999) 29; \\
O. Yasuda,   Acta. Phys. Polon. B {\bf 30} (1999) 3089; \\
M. Koike and J. Sato, Phys. Rev. D {\bf 61} (2000) 073012;
Erratum {\it ibid.} D {\bf 62} (2000) 079903.

\bibitem {t-matter}
S.~J.~Parke and T.~J.~Weiler,
Phys.\ Lett.\ B {\bf 501} (2001) 106.


\bibitem{MNjhep01}
H. Minakata and H. Nunokawa, JHEP {\bf 0110} (2001) 001 [hep-ph/0108085].

%\bibitem{CHOOZ}
%CHOOZ collaboration, M. Apollonio {\it et al.}, 
%Phys. Lett. B {\bf 466} 415. 

\bibitem {KTY02}
K.~Kimura, A.~Takamura and H.~Yokomakura, hep-ph/0203099.

\bibitem {N92}
V.~N.~Naumov, Int. J. Mod. Phys. D {\bf 1} (1992) 379.

\bibitem {HS00}
P.~F.~Harrison and W.~G.~Scott, Phys. Lett. B {\bf 476} (2000) 349.

\bibitem {ZS88}
H.~W.~Zaglauer and K.~H.~Schwarzer, Z.\ Phys.\ C {\bf 40} (1988) 273.
%%CITATION = ZEPYA,C40,273;%%


\bibitem {nufact}
A.~Cervera, A.~Donini, M.~B.~Gavela, J.~J.~Gomez Cadenas,
P.~Hernandez, O.~Mena and S.~Rigolin,
Nucl.\ Phys. \ B {\bf 579} (2000) 17
[{\it Erratum-ibid.}\ B {\bf 593} (2000) 731]; \\
C.~Albright {\it et al.}, hep-ex/0008064; \\
V.~Barger, S.~Geer, R.~Raja and K.~Whisnant,
Phys.\ Rev.\ D {\bf 63} (2001) 113011; \\
J.~Pinney and O.~Yasuda,
Phys.\ Rev.\ D {\bf 64} (2001) 093008.
%%CITATION = HEP-PH 0105087;%%


\bibitem{taup2001}
H. Minakata and H. Nunokawa,
Talk presented at 7th International Workshop on Topics in
Astroparticle and Underground Physics (TAUP2001),
Laboratori Nazionali del Gran Sasso, Italy, September 8-12, 2001, 
hep-ph/0111131, to appear in Nucl. Phys. Proc. Suppl.



\bibitem {BurguetC}
J. Burguet-Castell, M.B. Gavela, J.J. Gomez-Cadenas, P. Hernandez
and O. Mena,
Nucl. Phys. B {\bf 608} (2001) 301.


%\bibitem {KMN02}
%T.~Kajita, H.~Minakata, and H.~Nunokawa,
%Phys. Lett. B {\bf 528} (2002) 245.[hep-ph/0112345]

\bibitem {MNP02}
H.~Minakata, H.~Nunokawa, and S.~J.~Parke, work in progress.





\end{thebibliography}
\end{document}